\documentclass[aps,pra,onecolumn,nofootinbib,longbibliography,a4paper,superscriptaddress,11pt,tightenlines,accepted=2021-02-16]{quantumarticle}
\pdfoutput=1
\usepackage[utf8]{inputenc}
\usepackage[english]{babel}
\usepackage[T1]{fontenc}
\usepackage{amsmath}
\usepackage{hyperref}

\usepackage{tikz}
\usepackage{lipsum}

\usepackage{amsfonts}

\usepackage{amsthm}
\usepackage{amscd}
\usepackage{mathtools}
\usepackage{t1enc}
\usepackage[mathscr]{eucal}
\usepackage{indentfirst}
\usepackage{graphicx}
\usepackage{graphics}
\usepackage{pict2e}
\usepackage{epic}
\numberwithin{equation}{section}
\usepackage[margin=2.9cm]{geometry}
\usepackage{epstopdf} 
\usepackage{physics}

\usetikzlibrary{arrows}
\usetikzlibrary{petri}
\usetikzlibrary{graphs,graphs.standard,quotes,patterns}

\renewcommand\thesection{\arabic{section}}

\theoremstyle{plain}\label{key}
\newtheorem{Theorem}{Theorem}
\newtheorem{Lemma}[Theorem]{Lemma}

\newtheorem{Proposition}[Theorem]{Proposition}

\theoremstyle{definition}
\newtheorem{Definition}{Definition}

\newtheorem{?}[Definition]{Problem}
\newtheorem{Example}[Definition]{Example}

\DeclareMathOperator{\sgn}{sign}
\DeclareMathOperator{\Real}{Re}
\DeclareMathOperator{\Imag}{Im}

\newcommand{\Cd}{\ensuremath{\mathbb{C}^d}}

\newcommand{\Rd}{\ensuremath{\mathbb{R}^d}}

\newcommand{\C}{\ensuremath{\mathbb{C}}}
\newcommand{\M}{\ensuremath{\mathcal{M}}}

\newcommand{\R}{\ensuremath{\mathbb{R}}}

\newcommand{\Stab}{\ensuremath{\textup{STAB}_n}}

\newcommand{\D}{\ensuremath{\mathcal{D}}}
\newcommand{\Dn}{\ensuremath{\D_n}}

\newcommand{\st}{\ensuremath{\textup{s.t.}}}

\def\zz{\mathbb{Z}}
\def\cc{\mathbb{C}}
\def\rr{\mathbb{R}}

\begin{document}

\title{Stabilizer extent is not multiplicative}

\author{Arne Heimendahl}
\affiliation{Department Mathematik/Informatik, Universit\"at zu K\"oln, Weyertal 86--90, 50931 Cologne, Germany}
\email{arne.heimendahl@uni-koeln.de}
\author{Felipe Montealegre-Mora}
\affiliation{Institute for Theoretical Physics, Universit\"at zu K\"oln, Z\"ulpicher Str.\ 77, 50937 Cologne, Germany}
\author{Frank Vallentin}
\affiliation{Department Mathematik/Informatik, Universit\"at zu K\"oln, Weyertal 86--90, 50931 Cologne, Germany}
\author{David Gross}
\affiliation{Institute for Theoretical Physics, Universit\"at zu K\"oln, Z\"ulpicher Str.\ 77, 50937 Cologne, Germany}

\maketitle

\begin{abstract}
  The Gottesman-Knill theorem states that a Clifford circuit acting on stabilizer states can be simulated efficiently on a classical computer.
  Recently, this result has been generalized to cover inputs that are close to a coherent superposition of polynomially many stabilizer states.
  The runtime of the classical simulation is governed by the \emph{stabilizer extent}, which
  roughly measures how many stabilizer states are needed to approximate the state.
  An important open problem is to decide whether the extent is multiplicative under tensor products.
  An affirmative answer would yield an efficient algorithm for computing the extent of product inputs, while a negative result implies the existence of more efficient classical algorithms for simulating large-scale quantum circuits.
  Here, we answer this question in the negative.
  Our result follows from very general properties of the set of stabilizer states, such as having a size that scales subexponentially in the dimension, and can thus be readily adapted to similar constructions for other resource theories.
\end{abstract}

	\section{Introduction and Summary of results}
In the model of quantum computation with magic states \cite{Bravyi2004}, stabilizer circuits, whose computational power is limited by the Gottesmann-Knill theorem \cite{Gottesman1998,Aaronson2004}, are promoted to universality by implementing non-Clifford gates via the injection of \emph{magic states}. 
There has been a long line of research with the goal of designing classical algorithms to simulate such circuits.

\emph{Quasiprobability-based methods} \cite{Pashayan2015,BermejoVega2016,Campbell2017,Frembs2018,Raussendorf2019,Heinrich2018} work on the level of density operators. 
The starting point is the observation that the (qudit) Wigner function \cite{Gross2006} of stabilizer states is given by a probability distribution on phase space and thus gives rise to a classical model.
Similar to the \emph{quantum Monte-Carlo} method of many-body physics, one can then devise randomized simulation algorithms whose runtime scales with an appropriate ``measure of negativity'' of more general input states.

\emph{Stabilizer rank methods} \cite{Bravyi2015,Seddon2020,Bravyi2016}, 
on the other hand, work with vectors in Hilbert space.
The idea is to expand general input vectors as a coherent superposition of stabilizer states.  
The smallest number of stabilizer states required to express a given vector in this way is its \emph{stabilizer rank}.
Bravyi, Smith, and Smolin \cite{Bravyi2015} proposed a fast simulation algorithm. 
Its time complexity scales with the stabilizer rank rather than the -- often much higher -- dimension of the Hilbert space.
Bravyi and Gosset \cite{Bravyi2016} generalized this procedure to cover \emph{approximate stabilizer decompositions}.

No efficient methods are known for computing the stabilizer rank analytically or numerically.
To address this issue, Bravyi \emph{et al.} \cite{Bravyi2018}
introduced a computationally better-behaved convex relaxation:
the \emph{stabilizer extent} (see Definition~\ref{Def:ComplexExtent}).
The central \emph{sparsification lemma} of \cite{Bravyi2018} states that a stabilizer decomposition with small extent can be transformed into a sparse decomposition that is close to the original state. 
In this way, the stabilizer extent defines an operational measure for the degree of ``non-stabilizerness''.
We work in a slightly more general setting than~\cite{Bravyi2018}, where the role of the
stabilizer states is replaced by a finite set $\D\subset\Cd$
which spans $\Cd$, referred to as a \emph{dictionary}.

\begin{Definition}[\cite{Bravyi2018}]\label{Def:ComplexExtent}
	Let $ \D \subset \Cd $ be a finite set of vectors spanning $\Cd$. 
	For an element $ x \in \Cd $, the \emph{extent} of $ x $ with respect to $ \D $ is defined as
	\begin{align*}
		\xi_{\D}(x) =\min  \left \{ \norm{c}_1^2 \, : \,  c \in \C^{|\D|}, \, x = \sum_{s \in \D } c_s s \,   \right \},
	\end{align*} 
	where $ \norm{c}_1 = \sum_{s \in \D } |c_s| $. If $d=2^n$ and $\D=\Stab$ is the set 
	of stabilizer states, then $\xi_\D(x)$ is the \emph{stabilizer extent} of $x$, and the 
	notation is shortened to $\xi(x)$.
\end{Definition}
As is widely known, $ \ell_1 $-minimizations such as $ \xi_{\D} $ can be formulated as convex optimization problems (see for example \cite{Boyd2004}). 
In the complex case this is a \emph{second order cone problem} \cite{Alizadeh2003}, whose complexity scales polynomially in  $ \max(d,|\D|) $. 
In particular, the complexity of determining the stabilizer extent of an arbitrary vector, 
$\xi(x)$, scales exponentially in the number of qubits. 
Thus, the question arises whether it is possible to simplify the computation of $ \xi_{\D} $ for certain inputs, e.g.\ product states of the form $ \psi = \otimes_j \psi_j $.

Since the set of stabilizer states is closed under taking tensor products, one can easily see that the stabilizer extent is submultiplicative, that is 
$ \xi(\otimes_{j}\psi_j) \le \prod_{j}\xi(\psi_j) $ for any input state $ \otimes_{j}\psi_j $.
Bravyi \emph{et al.} proved that it is actually multiplicative if the factors 
are composed of $ 1 $-, $2 $- or $ 3 $-qubit states. 

Our main result is that stabilizer extent is \emph{not} multiplicative in general.
In fact, our result does not depend on the detailed structure of stabilizer states, but holds for fairly general families of dictionaries.
The properties used --- prime among them that the size of the dictionaries scales subexponentially with the Hilbert space dimension --- are listed as Properties (i) to (v) in the following theorem.

\begin{Theorem}\label{maintheorem}
	Let $ (\Dn) $ be a sequence of dictionaries with $  \Dn \subset  (\C^{d_0})^{\otimes n}$ and $ \D_1 \subset \C^{d_0} $ for some fixed integer $ d_0 $. Assume that $ (\Dn) $ satisfies the following properties: 
	\begin{enumerate}
		\item[(i)\label{itm:1}]  
		Normalization: 
		$ \langle s, s\rangle = 1 $ for all $ s \in \Dn $.
		\item[(ii)\label{itm:2}]  
		Subexponential size:
		\begin{align*}
			\log_{d_0}|\Dn| \le o\left (\sqrt{d_0^n}\right ).
		\end{align*}
		\item[(iii)\label{itm:3}] 
		Closed under complex conjugation: if $ s \in \Dn $, then $ s^* \in \Dn$. 
		\item[(iv)\label{itm:4}] 
		Closed under taking tensor products:
		\begin{align*}
			\D_{n_1} \otimes \D_{n_2}:= \{ s_1 \otimes  s_2 \, : \,s_1 \in \D_{n_1}, \,  s_2 \in \D_{n_2}  \} \subset \D_{n_1+n_2}. 
		\end{align*}
		\item[(v)\label{itm:5}] 
		Contains the maximally entangled state: 
		For every $ n $, the maximally entangled state 
		\begin{align*}
			\Phi = \frac{1}{\sqrt{d_0^n}}\sum_{k \in \zz_{d_0}^n} e_k \otimes e_k \in \D_{2n}
		\end{align*}
		is contained in the dictionary $\D_{2n}$.
		Here, $\{e_k\}$ is the standard (``computational'') basis of $(\cc^{d_0})^{\otimes n}$.
	\end{enumerate}
	Let $ \psi  \in (\C^{d_0})^{\otimes n}$ be a unit vector. Then
	\begin{align*}
		\textup{Pr} [\xi_{\D_{2n}}(\psi \otimes \psi^*) < \xi_{\Dn}(\psi) \xi_{\Dn}(\psi^*) ] \ge 1 - o(1).
	\end{align*}
	In particular, for sufficiently large $ n $, the extent with respect to the dictionary sequence $ (\Dn) $ is \emph{strictly} submultiplicative.
\end{Theorem}
Parts of the proof of Theorem \ref{maintheorem} follow the proof of non-multiplicativity of the \emph{stabilizer fidelity} \cite[Lemma 10]{Bravyi2018}. As a crucial extra ingredient, we carefully analyze the dual second order cone formulation of the extent and exploit \emph{complementary slackness} to prove the fact that the optimal \emph{dual witness} is generically unique. 

Note that the main theorem also implies that other \emph{magic monotones} recently defined in \cite{Seddon2020} (\emph{mixed state extent, dyadic negativity,} and \emph{generalized robustness}) fail to be multiplicative, since they all coincide with the stabilizer extent on pure states \cite{Regula2017}.

The remaining part of paper is organized as follows: 
In Section~\ref{Section:ProofStrategy}, we outline the geometric intuition behind the argument.
The rigorous proof is given in Section~\ref{SectionComplexCase}.
As an auxiliary result, we present an optimality condition on stabilizer extent decompositions in Section~\ref{SectionOptimalityConditions}.

\section{Proof strategy}\label{Section:ProofStrategy}

In this section, we explain the geometric intuition behind the main result.
To simplify the exposition, we present a version of the argument for real vector spaces.

\begin{figure}[h]
	\centering
	\begin{tikzpicture}
		\draw (0,0) circle (2cm);
		\draw (-2,-3) -- (-2,3);
		\draw (-3,-2) -- (3,-2);
		\draw (-3,2) -- (3,2);
		\draw (2,-3) -- (2,3);
		\draw[->] (0,0) -- (0,3);
		\draw[->] (0,0) -- (3,0);
		
		\node[inner sep=0, minimum size=4cm,fill=blue!80, opacity=0.5] at (0,0) {};
		
		\draw[-] (2.2,1) -- (3,1.4);
		\node[color=red!80] at (3.5,1.4) {$C_{y_2}$};
		
		\draw[-] (1.8,-1.5) -- (3,-1.1);
		\node[color=blue!80] at (3.5,-1.1) {$M_{\D}$};
		
		\draw[->, color=black] (0,0) -- (0,3);
		\draw[->, color=black] (0,0) -- (3,0);
		\fill[radius=1pt]
		(0,0) circle [];
		\node[inner sep=1, minimum size=2.48cm, fill=red!80, opacity=.5] at (1.25,1.25) {};
		
		\fill[radius=1pt]
		(-2,0) circle[] node[left] {$-s_1$}
		(-2,2) circle[] node[above right] {$y_1$}
		(0,2) circle[] node[above left] {$s_2$}
		(2,2) circle[] node[above right] {$y_2$}
		(2,0) circle[] node[below right] {$s_1$}
		(2,-2) circle[] node[below right] {$-y_1$}
		(0,-2) circle[] node[below left] {$-s_2$}
		(-2,-2) circle[] node[below left] {$-y_2$}
		(0,0) circle [];
	\end{tikzpicture}
	\caption{The polytope $ M_\D $ for the dictionary $ \D = \{s_1,s_2\} \subset \mathbb{S}^1$ and the normal cone $ C_{y_2} $ of the vertex $ y_2 $. The active inequalities at $ y_2 $ yield the extreme rays of $ C_{y_2} $. \label{figure:NormalCone}}
\end{figure}
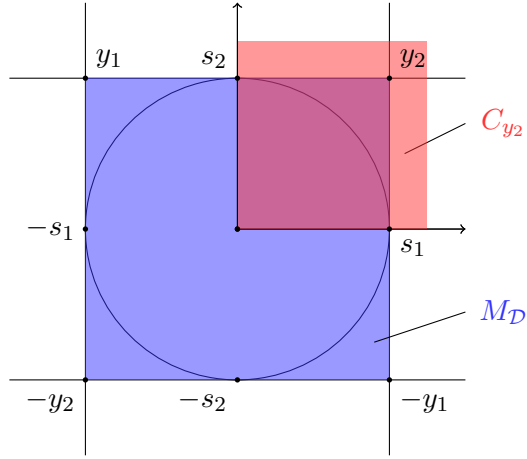

We recall the convex geometry underlying the problem. 
In the real case, the extent can be formalized as a \emph{basis pursuit problem}:
\[
\begin{array}{lll}
	\sqrt{\xi_{\D}(x)} \; = \; & \min \; & \displaystyle\sum\limits_{s \in \D }|c_s|   \\[0.5ex]
	&\st & c_s \in \R \  (s \in \D), \\[0.5ex]
	& &\displaystyle\sum_{s \in \D }c_s s = x.
\end{array}
\]
This type of optimization can be formulated as linear program (see e.g. \cite[Chapter 6]{Boyd2004}). Using standard techniques we can derive its dual form (see e.g.~\cite[Chapter 5]{Boyd2004}):
\[
\begin{array}{lll}
	\sqrt{\xi_{\D}(x)} 
	\; = \;
	& \max \; &x^\top y \\[0.5ex]
	&\st & y \in \Rd, \\[0.5ex]
	& &|s^\top y| \le 1 \textup{\ for all } s \in \D,
\end{array}
\]
where $ x^\top y := \sum\limits_{j= 1}^d x_j y_j $ denotes the inner product on $ \Rd $.
Let
\begin{align*}
	M_\D = \{ y \in \Rd \, : \,|s^\top y | \le 1 \textup{ for all } s \in \D \}
\end{align*}
be the region of feasible points for the dual program. 
Since $ \D $ is finite and contains a spanning set of $ \Rd $, the set $ M_\D $ is a polytope. 
The dual formulation implies that for each $x$, there exists a \emph{witness} $y$ among the vertices of $M_\D$ such that $\sqrt{\xi_\D(x)}=x^\top y$. 
Conversely, with each vertex $y\in M_\D$, one can associate the set of primal vectors $x$ for which $y$ is a witness:
\begin{align*}
	C_y = \left\{ x\in\Rd \,:\, \sqrt{\xi_\D(x) }= x^\top y \right\} 
	= \textup{cone} \left\{ (-1)^ks \, : \,s \in \D, \, k \in \{0,1\},  \, (-1)^ks^\top y = 1 \right\}.
\end{align*}
The cone over a set $ M $, denoted by $ \textup{cone}\{M \}$, is simply the set of all linear combinations with non-negative coefficients of a finite set of elements in $ M $.
It is easy to see that the $C_y$ are full-dimensional convex
cones that partition $\Rd$ as $y$ ranges over the vertices of
$M_\D$ (see Figure \ref{figure:NormalCone} for an
illustration). The cones $ C_y $ are called \emph{normal
	cones} and the induced partition of $ \Rd $ is referred to
as the \emph{normal fan} of $ M_\D $, see for example \cite{Ziegler1995}.
For $ x \in \Rd $, define the \emph{fidelity of $x$ with respect to $\D$} 
\begin{align*}
	\sqrt{F_\D(x) }:= \max_{s \in \D} | s^\top x |
\end{align*} 
as the maximal overlap of $ x $ with an element in $ \D $   
(the value $ \sqrt{F_\D(x)} $ can also be viewed as the $ \ell_\infty $-norm of $ x $ with respect to $ \D $).

These notions allow us to analyze how the extent of a vector $x$ changes when a word $w$ is added to the dictionary $\D$ (in the proof below, we will track the extent when the maximally entangled state is added to a product dictionary).
Indeed,
if $ x $ is contained in the interior of some $ C_y $,
and if $ |w^\top y | > 1  $, then the vertex $ y $ is infeasible for the dual program with respect to the dictionary $ \D \cup \{w\}$ (i.e., $ y \notin M_{\D \cup \{w\}} $), and therefore 
$ \xi_{\D \cup \{w\}}(x) < \xi_\D(x) $. 

Now, the argument of the proof of the main theorem proceeds in two steps:
\begin{enumerate}
	\item [(1)]  Assume $x$ is chosen Haar-randomly from the unit-sphere in $\rr^d$.
	Almost surely, there will be a \emph{unique} witness $y$, i.e.,\ $x$ will lie in the interior of some normal cone $C_y$ for some vertex $ y $ of  $ M_\D $.
	Moreover, the norm of $ y $ is large with high probability, $\|y\|_2^2 \approx O(d)$. 
	To see why the latter holds, note that
	\begin{align*}
		\|y\|_2^2 \geq (x^\top y)^2 = \xi(x) \geq \frac1{F_\D(x)},
	\end{align*}
	where the second inequality follows because $x/\sqrt{F_\D(x)} \in M_D$ is feasible for the dual (as realized in \cite{Bravyi2018}).
	A standard concentration-of-measure argument (as in \cite{Bravyi2018}, proof of Claim $ 2 $) shows that if $|\D|$ is not too large, the \emph{maximal} inner product-squared of $x$ with any element of $\D$ will be close to the expected inner product-squared with any \emph{fixed} unit vector $ v $, which is $ |x^\top v|^2 \approx 1/d $.
	\item [(2)]  Now consider $x\otimes x$. 
	With respect to the \emph{product dictionary} $\D\otimes \D$, one easily finds that $\xi_{\D \otimes \D}(x\otimes x)=\xi_\D(x) \xi_\D(x)$, and  that $y\otimes y$ is a unique witness and a vertex of  
	$ M_{\D \otimes \D} $.
	If $\Phi$ is the maximally entangled state,
	\begin{align*}
		\Phi^\top (y \otimes y) = d^{-1/2} \|y\|^2_2 = O(d^{1/2}) > 1.
	\end{align*}
	Thus adding $\Phi$ to the dictionary means that $y\otimes y$ becomes dually \emph{infeasible} (i.e., $ y \otimes y \notin M_{\D \otimes \D \cup \{\Phi\}} $).  It follows that the extent of $x\otimes x$ (in fact, the extent of \emph{any} element in the interior of $C_{y\otimes y}$) decreases if $\Phi$ is added.
\end{enumerate}

\section{Proof of the main theorem}\label{SectionComplexCase}

In preparation of proving the main theorem, we translate the convex geometry of $\ell_1$-minimization from the real case (sketched in the previous section) to the case of complex vector spaces.
This problem has been treated before in various places in the literature, including in \cite{Bravyi2015}, in the context of the theory of \emph{compressed sensing} (e.g.\ \cite{Rauhut}), and in greater generality in the convex optimization literature (e.g.\ \cite{Pataki2000}). 
As we are not aware of a reference that gives a concise account of all the statements required, we present self-contained proofs in Appendix~\ref{Appendix}.

We will use the superscripts $ R $ and $I$ to denote, respectively, the real and complex part of a vector. 
The extent then has the following dual formulation (c.f.\
Appendix~\ref{Appendix}):
\[
\begin{array}{lll}
	\sqrt{\xi_{\D}(\psi)} \; = \; & \max \; &  
	(\psi^R)^\top y^R+ (\psi^I)^\top y^I   \\[0.5ex]
	&\st &  y \in \Cd,			\\[0.5ex]
	&& \sqrt{F_{\D}(y)} \le 1,
\end{array}
\]
where 
\begin{align*}
	F_{\D}(y) = \max_{s \in \D}| \langle s,y \rangle  |^2 
\end{align*} 
and $ \langle s,y \rangle := \sum\limits_{j = 1}^d \overline{s_j} y_j $ denotes the inner product on $ \Cd $.

Let 
\begin{align*}
	M_\D = \{ y \in \Cd \, : \,|\langle s,y \rangle  | \le 1 \textup{ for all } s \in \D \}
\end{align*} be the set of feasible points for the dual. 
In contrast to the real case, $ M_\D $ is not a polytope, but $ M_\D $ is still a bounded convex set (viewed as a subset in $ \mathbb{R}^{2d} $ -- for a more detailed explanation, see Appendix~\ref{Appendix}).
Thus, by Krein-Millman,  $ M_\D $ is the convex hull of its
extreme points, which can be characterized as follows
(Appendix~\ref{Appendix} contains a proof):
\begin{Proposition}\label{Prop:ExtremePointsMD}
	A point $ y  \in M_\D$ is an extreme point of $ M_\D $ if and only if 
	\begin{align*}
		\big\{  s \in \D \, : \, |\langle s,y \rangle |= 1 \big\}
	\end{align*}
	is a spanning set for $\Cd$. 
\end{Proposition}

We will continue with an example of one extreme point of $ M_{\D} $ for $ \D = \textup{STAB}_n $ being the dictionary of $ n $-qubit stabilizer states.

\begin{Example}
	One extreme point for the set $\M_{\Stab}$ is the rescaled tensor-power $\psi_T^{\otimes n}/F(\psi_T^{\otimes n})$ of the magic $T$-state,
	\begin{align*}
		\psi_T := 
		\begin{pmatrix}
			\cos(\beta)\\
			e^{i\frac{\pi}{4}} \sin(\beta)
		\end{pmatrix},
	\end{align*}
	where $\beta = \frac{1}{2} \arccos(\frac{1}{\sqrt{3}})$.
	In this remark, we sketch why this is so.
	
	The vector $\psi_T^{\otimes n}$ satisfies $\xi(\psi_T^{\otimes n})=1/F(\psi_T^{\otimes n})$~\cite[Proposition 2]{Bravyi2018}.
	Now, $\psi_T\psi_T^\dagger=\frac13(\mathbb{I}+C+C^2)$ where $C$ is the Clifford matrix which cyclically permutes the Pauli matrices $\{X,Y,Z\}$. 
	This way, if $U = C^{i_1}\otimes\dots\otimes C^{i_n}$, then 
	\begin{align*}
		\langle U^\dagger s, \psi_T^{\otimes n}\rangle
		=
		\langle s, U\psi_T^{\otimes n}\rangle
		=
		\langle s, \psi_T^{\otimes n}\rangle \qquad \text{for all } s\in\Stab.
	\end{align*}
	It follows that the group generated by tensor products of $\{\mathbb{I},C\}$ acts on the optimizers of $F(\psi_T^{\otimes n})$.
	But the standard basis vector $e_0^{\otimes n}$ is one such optimizer~\cite[Lemma~2]{Bravyi2018} and
	\begin{align*}
		\mathrm{Span}\{e_0,\,Ce_0\}
		=
		\mathrm{Span}\{e_0,\,(e_0+e_1)/\sqrt{2}\}
		=
		\mathbb{C}^2.
	\end{align*}
	This shows that the optimizers of $ F(\psi_T^{\otimes n}) $ contain all tensor products of $ e_0 $ and $ (e_0+e_1)/\sqrt{2} $, which form a basis for $ (\C^2)^{\otimes n} $. 
	
	Finally, $\psi_T^{\otimes n}/F(\psi_T^{\otimes n})$ is an optimal dual witness for $\psi_T^{\otimes n}$.
	By Prop.~\ref{Prop:ExtremePointsMD}, then, this witness is extremal.
\end{Example}


Returning back to the general theory, we associate a \emph{normal cone} with every extreme point $y$: 
\begin{align}\label{def:NormalCone}
	%
	%
	C_y \; = \; &\left  \{ \psi \in \Cd 
	\,:\, 
	\langle \psi, y \rangle^R = \max_{p \in M_\D} \langle \psi,p \rangle^R\right \}  \nonumber  \\
	\; = \;
	& \textup{cone}\left\{  e^{i\phi} s  \, : \, 
	s\in \D, \phi\in\mathbb{R}, \,
	e^{i\phi} \langle s,y \rangle  = 1  
	\right \}. 
\end{align}
Notice that 
\begin{align*}
	\langle \psi, y \rangle^R = (\psi^R)^\top y^R + (\psi^I)^\top y^I.
\end{align*}

A final preparation step invokes \emph{complementary
	slackness} (Appendix \ref{Appendix} contains a proof):

\begin{Lemma}[Complementary slackness conditions]\label{Lemma:CS}
	Let $  y \in M_\D $ be \emph{any} optimal dual witness, i.e., $ \psi \in C_y $ and $ \sqrt{\xi_{\D}(\psi)}  = \langle \psi, y \rangle^R  $.
	Then for \emph{any} optimal extent decomposition $ \psi = \sum_{s \in \D } c_s s  $ with $ \sqrt{\xi_{\D}(\psi)}  = \sum_{s \in \D } |c_s|$  we have the following two conditions:
	\begin{enumerate}
		\item[(I)\label{itm:ComplementarySlackness}] If $ c_s \neq 0 $, then $ \langle s,y \rangle = c_s/|c_s|$.
		\item[(II)\label{itm:ComplementarySlackness2}] If $ |\langle s,y \rangle |< 1 $, then $c_s = 0$.
	\end{enumerate}
\end{Lemma}

The complementary slackness conditions have the following two consequences: 

First, assume that $ \psi = \sum_{s \in \D }c_s s $ is an optimal decomposition and that $ y \in \Cd$ optimal for the dual. 
From condition \hyperref[itm:ComplementarySlackness]{(I)}, we obtain
\begin{align*}
	|\langle \psi,y \rangle  | =\left|  \sum_{s \in \D }\overline{c_s} \langle s,y \rangle \right| 
	= \left|\sum_{s \in \D, c_s \neq 0 } \overline{c_s} \frac{c_s}{|c_s|} \right| 
	= \sum_{s \in \D} |c_s| 
	= \sqrt{\xi_{\D}(\psi)},
\end{align*}
so we can rewrite the dual program for the extent as
\begin{equation}
	\label{Eq:BravyiExtentFormulation}
	\begin{array}{lll}
		\xi_{\D}(\psi) \; = \; &\max \; & |\langle \psi,y \rangle |^2 \\[0.5ex]
		&\st & 	 y \in \Cd, \\[0.5ex]
		&&F_{\D}(y) \le 1,
	\end{array}
\end{equation}
which coincides with the dual formulation given in \cite{Bravyi2018}.
Since $\psi/\sqrt{F_\D(\psi)}$ is feasible for the dual, we get the natural lower bound  \cite{Bravyi2018}
\begin{align}\label{Eq:SelfwitnessLowerbound}
	\xi_{\D}(\psi) \ge \frac{1}{F_\D(\psi)}.
\end{align}

Secondly, if a state $ \psi $ is chosen Haar-randomly, the optimal
dual witness $ y $ for $ \xi_{\D}(\psi) $ is an extreme point \emph{and} unique 
of $ M_\D $ with probability one, because of the following observation: 
A generic $ \psi $ will not be contained in a proper subspace spanned by elements of $ \D $, since the finite collection of all these lower-dimensional subspaces has measure zero. 
Thus, generically, if we expand $ \psi = \sum_{s \in \D }c_s s$ in the dictionary $ \D $, the 
set $ \{  s \in \D \, : \,c_s \neq 0 \} $ has to span $ \Cd $. 

Now suppose we are given two optimal dual witnesses $ y_1,y_2 $ for $ \xi_{\D}(\psi) $.
Condition \hyperref[itm:ComplementarySlackness]{(I)}  of \ref{Lemma:CS} tells us that for \emph{all} optimal primal extent decompositions, both $ y_1 $ and $ y_2 $ are solutions of the system of linear equations: 
\begin{align*}
	\langle s,y \rangle = \frac{c_s}{|c_s|} \quad \textup{for all } c_s \neq 0.		
\end{align*}
However, this system has a unique solution because the words $ s \in \D$ with $ c_s \neq 0 $ span $ \C^d $ and  therefore, $ y_1 = y_2 $.
Such $ \psi$'s are also called \emph{non-degenerate} in convex optimization \cite{Alizadeh2003}.

Analogously to the case of a normal cone in a real-valued vector space, note that the interior $ \textup{int}(C_y) $ of a normal cone $ C_y $ consists of all points $ \psi $ whose dual witness is unique and the extreme point $ y $. 
This means that there exists an optimal extent decomposition
\begin{align*}
	\psi = \sum_{s \in \D} c_s s = \sum_{s \in \D} \alpha_s \, e^{i\phi_s} s,
\end{align*}
such that
\[
\alpha_s \ge 0, \; c_s = \alpha_s e^{i\phi s},\; e^{i\phi_s}
s \in C_y, \; \text{and } \{ s \in \D \, : \,c_s \neq 0 \}
\text{ spans } \Cd .
\]

With the above notion, we are able to describe how the extent is effected by adding a word $ w $ to the dictionary $ \D $.
As in the case of a real valued vector space, an extreme point $ y \in M_\D $ becomes dually infeasible if $ |\langle w,y \rangle  | > 1 $ (i.e., $ y \notin M_{\D \cup \{w\}} $). Hence, the extent of an element $ x $ decreases if $ y $ is the unique dual witness of $ x $, that is $ x \in \textup{int}(C_y) $. In summary, we get the following theorem:

\begin{Theorem}\label{Thm:AddingAWordComplexCase}
	Let $ \D \subset \Cd$ be a dictionary and let $ w \in \Cd $ with $ \langle w,w \rangle  = 1  $. Let $ \D^\prime = \D \cup \{ w \} $. Then,
	$ \xi_{\D^\prime}(x)  < \xi_{\D}(x) $, if and only if $ x \in \textup{int} (C_{{y}}) $ for an extreme point $ y \in M_\D$ with $ |\langle w,y \rangle  |> 1  $.
\end{Theorem}

In order to analyze the multiplicativity properties of the extent for product inputs, we now turn our attention to \emph{product dictionaries}. 
The argument starts with the observation that extreme points of 
$M_\D$ are 	closed under taking tensor products. That is, if $y_1,y_2$ are extreme points 
of dually feasible sets $ M_{\D_j} \subset \mathbb{C}^{d_j} $
for two dictionaries $\D_1$ and $\D_2$, 
then $ y_1 \otimes y_2$ is an extreme point of $M_{\D_1 \otimes \D_2}$, where
$\D_1 \otimes \D_2 \subset \mathbb{C}^{d_1} \otimes \mathbb{C}^{d_2} $ is the product
dictionary. 
Indeed, since $ y_1 \otimes y_2 \in M_{\D_1 \otimes \D_2} $ and the set
\begin{align*}
	&\left\{ s_1 \otimes s_2\in\D_1\otimes\D_2 
	\,:\, 
	|\langle s_1 \otimes s_2, y_1 \otimes y_2 \rangle |  = 1  \right\} \\
	\; = \;
	&\left\{ s_1 \otimes s_2\in\D_1\otimes \D_2
	\,:\, 
	|\langle s_j, y_j  \rangle  | = 1, \, j = 1,2  \right\}
\end{align*}
is a spanning set of $\mathbb{C}^{d_1} \otimes \mathbb{C}^{d_2}$. 
Moreover, by the characterization of the normal cone \eqref{def:NormalCone}, it follows immediately that the normal cone of $ y_1 \otimes y_2 $ has the form
\begin{align}\label{Eq_ProductNormalCone}
	C_{y_1 \otimes y_2} = \textup{cone} \{  e^{i\phi_{s_1}} s_1 \otimes e^{i\phi_{s_2}} s_2 \, : \,
	e^{i\phi_{s_j}} s_j \in C_{y_j}, \, j = 1,2   \}.
\end{align}

This allows us to derive the following multiplicativity property of product dictionaries: 
\begin{Lemma} \label{MultiplicativityForDictionaries}
	Consider two dictionaries $ \D_j \subset \mathbb{C}^{d_j} $ and extreme points $ y_j \in M_{\D_j} $, $ j =1,2 $. Then,
	$ C_{y_1} \otimes C_{y_2} \subset C_{y_1 \otimes y_2} $ and $ \textup{int}(C_{y_1}) \otimes \textup{int}(C_{y_2}) \subset \textup{int}(C_{y_1 \otimes y_2}) $.
	Therefore, 
	\begin{align*}
		\xi_{\D_1 \otimes \D_2}(\psi_1 \otimes \psi_2) = \xi_{\D_1 }(\psi_1)  \xi_{\D_2 }(\psi_2)
	\end{align*}
	for all $ \psi_j \in \mathbb{C}^{d_i} $.
\end{Lemma}

\begin{proof}
	We will prove $ C_{y_1} \otimes C_{y_2} \subset C_{y_1 \otimes y_2} $, the statement  $ \textup{int}(C_{y_1}) \otimes \textup{int}(C_{y_2}) \subset \textup{int}(C_{y_1 \otimes y_2}) $ can be proven analogously. 
	Let $ \psi_j \in C_j $, so
	\begin{align*}
		\psi_j = \sum_{s \in \D} \alpha_s^{j} \,  e^{i \phi_s^j} s, 
	\end{align*}
	where $ \alpha_s^j \ge 0 $ and if $ \alpha_s^j $ is positive, then $ e^{i\phi_s^j}s \in C_{y_j}$. 
	Thus,  
	\begin{align*}
		\psi_1 \otimes \psi_2 = \sum_{ s \otimes s^\prime \D_1 \otimes \D_2} \alpha_s^1 \alpha_{s^\prime}^2 \, (e^{i \phi_s^1 }  s \otimes  e^{i\phi_{s^\prime}^2} s^\prime) \in  C_{y_1 \otimes y_2}, 
	\end{align*}
	by Equation \eqref{Eq_ProductNormalCone}.
	
	In order to prove multiplicativity it suffices to
	observe that, by the definition of the normal cone and
	the extent formulation
	\eqref{Eq:BravyiExtentFormulation},
	\[
	\xi_{\D_1 \otimes \D_2}(\psi_1 \otimes \psi_2) 
	=
	| \langle \psi_1 \otimes \psi_2, y_1 \otimes y_2 \rangle  |^2 
	= 
	|\langle \psi_1, y_1 \rangle  |^2 \, |\langle \psi_2, y_2 \rangle  |^2   
	= \xi_{\D_1} (\psi_1)  \xi_{\D_2 }(\psi_2).  \qedhere
	\]
\end{proof}

Using the above lemma and the generic uniqueness of the dual witness $ y $, we are now able to prove our main theorem. 
We subdivide the proof in two parts, where the first part is an adaption of Claim $ 2 $ in \cite{Bravyi2018} to the class of dictionaries defined in Theorem~\ref{maintheorem}:
\begin{Proposition}\label{Prop:SmallOverlap}
	Assume that the dictionary sequence $( \D_n) $ with $ \D_n \subset (\C^{d_0})^{\otimes n} $ satisfies the assumptions of Theorem \ref{maintheorem}. Then, for a Haar-randomly chosen unit vector $ \psi \in (\C^{d_0})^{\otimes n} $ and some fixed $ \varepsilon > 0 $ it holds that
	\begin{align*}
		\textup{Pr}\left  [F_{\D_{n}}(\psi)  \le \frac{1}{\sqrt{d_0^n}+\varepsilon}  \right ]  \ge 1 - o(1).
	\end{align*}
	In particular, $ F_{\D_{n}}(\psi)  \le \frac{1}{\sqrt{d_0^n}+\varepsilon}  $ for sufficiently large $ n $ and a typical unit vector 
	$ \psi \in (\C^{d_0})^{\otimes n} $.
\end{Proposition}

\begin{proof}
	We fix a unit vector $ \omega \in (\C^{d_0})^{\otimes n} $ and choose a Haar-random unit vector $ \psi \in (\C^{d_0})^{\otimes n}$. Following the proof of Claim $ 2 $ in \cite{Bravyi2018} we can bound the probability of the event $ \{ |\langle \omega,\psi \rangle |^2 \ge x \} $ by 
	\begin{align*}
		\textup{Pr}[| \langle \omega,\psi \rangle |^2 \ge x  ]  = (1-x)^{d_0^n-1} \le e^{-x(d_0^n-1)}. 
	\end{align*}
	If we set $ x = (\sqrt{d_0^n}+\varepsilon)^{-1} $ for $ \varepsilon > 0 $ and use Properties \hyperref[itm:1]{(i)} 
	and \hyperref[itm:2]{(ii)}, 
	we can use a union bound to estimate the fidelity of $ \psi $ with respect to $ \Dn $ by 
	\begin{align*}
		\textup{Pr}\left[ \max_{s \in \D} |\langle\psi,s \rangle |^2 \ge \frac{1}{\sqrt{d_0^n}+\varepsilon}  \right]
		& \; \le \; 
		|\Dn|\cdot \exp\left( -\frac{d_0^n-1}{\sqrt{d_0^n}+ \varepsilon} \right) \\
		& \; \le \;
		\exp\left(  o\left(\sqrt{d_0^{n}}\right)\mathrm{ln}(d_0)  -\frac{d_0^n-1}{\sqrt{d_0^n}+ \varepsilon} \right),
	\end{align*}
	which converges to zero as $ n $ tends to infinity. 
\end{proof}

The proposition assures that randomly chosen unit vectors generically have small overlap with elements in the dictionary sequence. Starting from there, we proceed with the proof of the main theorem. 
\begin{proof}[Proof of Theorem \ref{maintheorem}]
	Let $ \psi  \in (\C^{d_0})^{\otimes n}$ be a unit vector satisfying $ F_{\Dn}(\psi) \le \frac{1}{\sqrt{d_0^n}+\varepsilon} $ for some $ \varepsilon > 0 $. Due to Proposition \ref{Prop:SmallOverlap}, this holds for a typical $ \psi $ and sufficiently large $ n $.
	As a consequence of \eqref{Eq:SelfwitnessLowerbound}, we can lower bound the extent of $ \psi $ by
	\begin{align*}
		\xi_{\D_{n}}(\psi) \ge \frac{1}{F_{\Dn}(\psi)} \ge \sqrt{d_0^n}+ \varepsilon.
	\end{align*}
	
	Let $ y \in M_{\Dn}$ be an optimal dual witness, so $ \psi \in C_y $. As pointed out earlier, we can further assume that $ y $ is an extreme point of $ M_{\Dn} $ and that $ y \in \textup{int}(C_y) $ generically. Applying Cauchy-Schwarz, we get a lower bound on the norm of $ y $ by 
	\begin{align}\label{WitnessBound}
		|\langle y,y \rangle | =|\langle y,y \rangle | \cdot |\langle\psi,\psi \rangle|  \ge | \langle \psi,y \rangle |^2  = \xi_{\D_{n}}(\psi) \ge \sqrt{d_0^n}+\varepsilon.
	\end{align} 
	
	Now consider $\psi \otimes \psi^*$. Assumption \hyperref[itm:3]{(iii)} 
	ensures that $ \xi_{\D}(\psi) = \xi_{\D}(\psi^*) $ and $ \psi^* \in \textup{int}(C_{y^*}) $.
	The proof of Lemma \ref{MultiplicativityForDictionaries} tells us that the extreme point $ y \otimes y ^*$ of $ M_{\Dn \otimes \Dn} $ is optimal for 
	\begin{align*}
		\xi_{\D_{n}\otimes \D_{n}}(\psi \otimes \psi^*) = \xi_{\Dn}(\psi) \xi_{\D_{n}}(\psi^*). 
	\end{align*}
	Moreover, it is the unique optimizer, as $ \psi \otimes \psi^* \in \textup{int}( C_y )\otimes \textup{int}(C_{y^*})  \subset \textup{int}(C_{y \otimes y^*}) $.
	
	Next, we add the maximally entangled state $\Phi$ to the dictionary and observe 
	\begin{align*}
		\xi_{\D_{n}\otimes \D_{n}}(\psi \otimes \psi^*) \ge \xi_{\D_{n}\otimes \D_{n} \cup \{\Phi\}}(\psi \otimes \psi^*),
	\end{align*}	
	since $\Dn\otimes\Dn \subset \Dn\otimes\Dn\cup\{\Phi\}$.  
	%
	The norm estimation \eqref{WitnessBound} of $ y $ yields
	\begin{align*}
		\max_{s \in \D_n \otimes \Dn \cup \{\Phi\}} |\langle s, y \otimes y^* \rangle  |^2 
		&\ge
		| \langle \Phi,  y \otimes y^* \rangle |^2 
		=
		\Big |\frac{1}{\sqrt{d}} \sum_{k \in \zz_{d_0}^n}\langle y,e_k\rangle  \langle y^*,e_k\rangle \Big |^2 \\
		&= \frac{1}{d} |\langle y,y \rangle |^2   > 1,
	\end{align*}
	therefore $ y \otimes y^* $ is not contained in the set of dually feasible points $ M_{\Dn \otimes \Dn \cup \{\phi\}} $ of the dictionary $ \Dn \otimes \Dn \cup \{\phi\} $.
	Since $ y \otimes y^* \in \textup{int}(C_{y \otimes y^*})$ we can apply Theorem \ref{Thm:AddingAWordComplexCase} to obtain 
	$\xi_{\D_{n}\otimes \D_{n} \cup \{\Phi\}}(\psi \otimes \psi^*) < \xi_{\D_{n}\otimes \D_{n}}(\psi \otimes \psi^*)$.
	
	To conclude, because of \hyperref[itm:4]{(iv)} and
	\hyperref[itm:5]{(v)},
	\[
	\xi_{\D_{2n}}(\psi \otimes \psi^*) 
	\leq 
	\xi_{\D_{n}\otimes \D_{n} \cup \{\Phi\}}(\psi \otimes \psi^*)
	<
	\xi_{\D_{n}\otimes \D_{n}}(\psi \otimes \psi^*)
	=
	\xi_{\D_n}(\psi)\xi_{\D_n}(\psi^*),
	\]              
	which proves the desired result.
\end{proof}

\section{An optimality condition for the stabilizer extent}\label{SectionOptimalityConditions}
In this section we fix the dictionary sequence to be the set of $ n $-qubit stabilizer states $ \Stab $ and we will derive a condition on optimal stabilizer extent decompositions. 
(While preparing this document, we learned that this fact had already been observed earlier \cite{Campbell2020}, but it does not seem to be published).

Let $ P_n = \big \{  \bigotimes_{i=1}^n W_i \, : \,W_i \in \{ I,X,Y,Z \} \big \} $ be the set of $ n $-qubit Pauli matrices. 
The set of stabilizer states can be decomposed in a disjoint union of orthonormal bases, where each basis is labeled by a maximally commuting set $ \mathcal{S} \subset P_n$ of Pauli matrices (see \cite{NielsenChuang2011}, Chapter 10, or \cite{Gross2006,Kueng2015} for details). The projectors on the basis elements can be written as 
$ ss^\dagger= \frac{1}{2^n}\sum_{\sigma \in \mathcal{S} } (-1)^{k_\sigma} \sigma   $, where $ k_\sigma \in \{0,1\} $ has to be chosen in a way such that $ \{  (-1)^{k_\sigma}\sigma \, : \,\sigma\in\mathcal{S}\}$ is a closed matrix group with $ 2^n $ elements.

\begin{Theorem}
	Let $ \psi $ be an $ n $-qubit state. Suppose that $ \psi = \sum c_s s $ is an optimal stabilizer extent decomposition, that is $ \xi(\psi) = \Big (\sum_{s \in \D } |c_s| \Big )^2 $. Then there is at most one non-zero $ c_s $ for the words $ s$ that are labeled by the same orthonormal basis. 
\end{Theorem}
For the proof of the theorem, we will make use of the \emph{Clifford group} $ \mathcal{C}_n$. For our purpose this is the unitary group that preserves the set $ \Stab $, i.e., if $ U \in \mathcal{C}_n $, then $ Us \in \Stab $ for all $ s \in \Stab $ (more details can be found in \cite{Gross2006}). 
\begin{proof}
	First, we prove the statement for the $ 1$-qubit case. 
	The $ 1 $-qubit stabilizer dictionary is  given by the disjoint union of three orthonormal bases 
	\begin{align*}
		\textup{STAB}_1 = \mathcal{B}_1 \, \dot{\cup} \,  \mathcal{B}_2 \,  \dot{\cup} \, \mathcal{B}_3,
	\end{align*}
	where the three orthonormal stabilizer bases are given by 
	\begin{align*}
		\mathcal{B}_1 =\left    \{ \begin{pmatrix}
			1 \\ 0
		\end{pmatrix},\begin{pmatrix}
			0 \\ 1 
		\end{pmatrix}  \right  \}, \quad  \mathcal{B}_2=\left  \{   \frac{1}{\sqrt{2}} \begin{pmatrix}
			1 \\ i
		\end{pmatrix},  \frac{1}{\sqrt{2}}\begin{pmatrix}
			1 \\ -i
		\end{pmatrix}  \right \}, \quad  \mathcal{B}_3=  \left \{  \frac{1}{\sqrt{2}}\begin{pmatrix}
			1 \\ 1
		\end{pmatrix}, \frac{1}{\sqrt{2}}\begin{pmatrix}
			1 \\ -1 
		\end{pmatrix} \right \}. 
	\end{align*}
	Because the Clifford group acts transitively on $\{\mathcal{B}_1,\,\mathcal{B}_2,\,\mathcal{B}_3\}$ and maps optimal decompositions to optimal decompositions -- i.e. if $ \psi = \sum_{s \in \Stab} c_s s $ is optimal, then so is $ U\psi = \sum_{s \in \Stab} c_s (Us)$ -- it suffices to prove the statement for a single basis, e.g.\ $\mathcal{B}_1$.

	%
	%
	
	So suppose that we have decomposition of some state $ \psi = \sum_{s \in \text{STAB}_1} c_s s  $ with non-negative coefficients in the basis $ \mathcal{B}_1 $.
	Since optimal $ \ell_1 $-decompositions are invariant under scaling with a complex number, we may assume that the part of the decomposition realized by $ \mathcal{B}_1 $ is of the form
	\begin{align*}
		\omega = 1\begin{pmatrix}
			1 \\ 0
		\end{pmatrix} + z \begin{pmatrix}
			0 \\ 1 
		\end{pmatrix}, \quad \text{or} \quad \omega = z\begin{pmatrix}
			1 \\ 0
		\end{pmatrix} + 1 \begin{pmatrix}
			0 \\ 1 
		\end{pmatrix}
	\end{align*}
	with $ z = x+iy \in \C $ and $ |x|+|y| \le 1 $.
	Hence, the coefficients have $ \ell_1 $-norm $ 1+|z| = 1+ \sqrt{x^2+y^2}$.
	If $ \omega $ is of the first form, then we can also decompose it as   
	\begin{align}\label{eq:newDecompOmega}
		\omega = (\sqrt{2}|x|) \cdot  \frac{1}{\sqrt{2}} \begin{pmatrix}
			1 \\ \sgn(x)
		\end{pmatrix}+ (\sqrt{2}|y|) \cdot \frac{1}{\sqrt{2}} \begin{pmatrix}
			1 \\ \sgn(y)i
		\end{pmatrix}+ (1-|x|-|y|) \cdot \begin{pmatrix}
			1 \\ 0
		\end{pmatrix}
	\end{align}
	and the $ \ell_1$-norm of the coefficients in this decomposition is 
	\begin{align*}
		\sqrt{2}|x| + \sqrt{2}|y| + (1-|x|-|y|) = 1 + (\sqrt{2} - 1) (|x|+|y|)  < 1+ \frac{1}{2} (|x|+|y|).
	\end{align*}
	But 
	\begin{align*}
		\sqrt{(\xi(\omega)) }\le 1+ \frac{1}{2}(|x|+|y|) < 1+\sqrt{x^2+y^2} = 1 + |z|, 
	\end{align*}
	so the decomposition of $ \omega $ using only elements of $ \mathcal{B}_1 $ is not optimal. By changing the two coordinates,  we can argue analogously if $ \omega = \begin{pmatrix}
		z \\ 1
	\end{pmatrix}$. 
	Updating the decomposition of $ \psi  $ to $ \psi = \sum_{s \in \text{STAB}_1} \hat{c}_s s$ via the new decomposition of $ \omega $ \eqref{eq:newDecompOmega}, we also get a new decomposition of $ \psi $ with lower $ \ell_1 $-norm and only one non-zero coefficient for the basis $ \mathcal{B}_1 $. 
	This follows by comparing $ \sum_{s \in \text{STAB}_1} |c_s| $ with $ \sum_{s \in \text{STAB}_1} |\hat{c}_s| $ via the triangle inequality. 
	
	For the $ n $-qubit case assume that $ \psi = \sum c_s s $ is a stabilizer decomposition with $ c_sc_{s^\prime} \neq 0 $ for two stabilizer states $ s,s^\prime \in \Stab $ belonging to the same orthonormal basis. Due to invariance of $ \xi $ under the Clifford group and its transitive action on orthonormal stabilizer bases, we may choose any orthonormal stabilizer basis. By possibly applying another Clifford unitary, we may even assume that 
	$ s= e_0 \otimes  e_0 \cdots e_0,\, s^\prime = e_1 \otimes e_0 \cdots e_0$.
	But if we consider the decomposition of the unnormalized state 
	\begin{align*}
		\omega = 
		c_se_0 \otimes  e_0 \otimes\cdots\otimes e_0
		+ 
		c_{s^\prime}  e_1 \otimes e_0\otimes\cdots\otimes e_0 
		= 
		(c_s e_0+ c_{s^\prime}e_1) \otimes e_0\otimes\cdots\otimes e_0,
	\end{align*} 
	the $ 1 $-qubit case result together with the fact that stabilizer states are closed under taking tensor products can be applied to see that the decomposition of $ \omega $ is not optimal. 
	Now, the crucial observation is that if $ \psi = \sum c_s s $ is an optimal stabilizer extent decomposition, then $\omega = c_s s + c_{s^\prime} s^\prime $ is an optimal decomposition for $ \omega $. But as the decomposition of $ \omega $ is not optimal, neither is the one of $ \psi $.
\end{proof}

There is an interesting connection between the derived optimality condition and the geometric properties of the \emph{stabilizer polytope} $ SP_n $, which is the convex hull of the projectors onto stabilizer states, i.e., $ SP_n = \textup{conv}\{ ss^\dagger \, : \,s \in \Stab\} $.
As shown in \cite{Heimendahl2019,Epstein2015}, two stabilizer projectors are connected by an edge if and only if they do not belong to the same orthonormal stabilizer basis. Thus, we can reformulate the above result: \\
If $ \psi = \sum c_s s$ is an optimal stabilizer extent decomposition and $ c_s c_{s^\prime} \neq 0 $, then the set $\textup{conv} \{ ss^\dagger, \, s^\prime (s^\prime)^\dagger  \}$ is an edge of $ SP_n $.

\section*{Summary and outlook}

We have settled an open problem in stabilizer resource theory,
by showing that the stabilizer extent is generically
sub-multiplicative in high dimensions.  What is striking is
that the previous multiplicativity results for one to three
qubit states \cite{Bravyi2015} made use of the detailed
structure of the set of stabilizer states.  In
contrast, our counterexample involves only a small number of
high-level properties of the stabilizer dictionary.  Therefore,
we see this work as evidence that $\ell_1$-based
complexity measures on tensor product spaces should be
expected to be strictly sub-multiplicative in the absence of
compelling reasons to believe otherwise.  In particular, it
seems highly plausible that the assumptions that go into
Theorem~\ref{maintheorem} can be considerably weakened.  We leave this problem
open for future analysis.

\section*{Acknowledgments}

We thank Markus Heinrich, Earl Campbell, Richard K\"ung, and James Seddon for interesting discussions and feedback.

This work has been supported by the DFG (SPP1798 CoSIP),
Germany's Excellence Strategy -- Cluster of Excellence
\emph{Matter and Light for Quantum Computing} (ML4Q)
EXC2004/1, Cologne's Key Profile Area \emph{Quantum Matter and
	Materials}, and the European Union's Horizon 2020 research
and innovation programme under the Marie Sk\l{}odowska-Curie
agreement No 764759. The third named author is partially
supported by the SFB/TRR 191 ``Symplectic Structures in
Geometry, Algebra and Dynamics'' and by the project ``Spectral
bounds in extremal discrete geometry'' (project number
414898050), both funded by the DFG.

\appendix

\section{Formulating the extent as a second order cone
	program}\label{Appendix}

Here, we write the extent of Definition \ref{Def:ComplexExtent} with respect to a complex dictionary $ \D \subset \Cd $ as a real second order cone program in standard form \cite{Alizadeh2003}. We impose the condition that the elements in $ \D $ are normalized, i.e., $ \langle w,w \rangle = 1 $.
For an optimal decomposition $ \psi = \sum_{s \in \D } c_s s $ we set $ c^R_s = \Real c_s $ and $ c^I_s = \Imag c_s $. 
The standard primal version of the extent is given by

\[
\begin{array}{lll}
	\sqrt{\xi_{\D}(\psi)} \; = \; &\min \; &
	\displaystyle \sum_{s \in \D } t_s   \\[0.5ex]
	&\st &\displaystyle \sum_{s \in \D} 
	\begin{bmatrix}
		s^R & -s^I  & 0 \\ s^I & s^R  & 0 
	\end{bmatrix} \cdot 
	\begin{bmatrix}
		c_s^R  \\ c_s^I \\ t_s 
	\end{bmatrix}  = \begin{bmatrix}
		\psi^R \\ \psi^I 
	\end{bmatrix}   \\[4ex]
	& &  (c_s^R,c_s^I,t_s) \in \mathcal{L}^{2+1} \, (s \in \D),
\end{array}
\]
where
\[
\mathcal{L}^{2+1} = \left \{ (x_1,x_2,t) \in \mathbb{R}^3
\, : \,\sqrt{x_1^2+x_2^2} \le t \right  \}
\]
is the $ 3 $-dimensional Lorentz cone. 
As the program is in primal standard form, we can derive its
dual formulation:
\begin{equation}
	\label{FirstDualConstraint}
	\begin{array}{ll}
		\max  \; &  (\psi^R)^\top y^R + (\psi^I)^\top y^I \\[0.5ex]
		\st &  \begin{bmatrix}
			(s^R)^\top & (s^I)^\top \\ (-s^I)^\top & (s^R)^\top  \\ 0 & 0 
		\end{bmatrix} \cdot 
		\begin{bmatrix}
			y^R \\ y^I 
		\end{bmatrix}  + z_s = \begin{bmatrix}
			0 \\ 0 \\ 1
		\end{bmatrix}  \textup{ for all } s \in \D,  	\\[4ex]
		&z_s \in \mathcal{L}^{2+1} \, (s \in \D), \,  (y^R,y^I) \in \mathbb{R}^{2d}.
	\end{array}
\end{equation}
Since $ \D $ contains a basis of $ \Cd $, both programs are strictly feasible and strong duality holds, so the optimal values for $ \min $ and $ \max $ coincide. 
The dual constraints are equivalent to $ \max_{s \in \D} |
\langle s,y \rangle  | \le 1 $, where $ y = y^R +iy^I \in \Cd
$. Thus, we can rewrite the dual as
\[
\begin{array}{ll}
	\max \; & (\psi^R)^\top y^R + (\psi^I)^\top y^I  \\[0.5ex]
	\st & |\langle s,y \rangle  |\le 1 \textup{ for all } s \in \D,			\\[0.5ex]
	& y \in \Cd.	
\end{array}
\]
Next, we prove Proposition \ref{Prop:ExtremePointsMD}, which gives a characterization of the extreme points of the set of dually feasible points 
$ M_\D = \{  y \in \Cd \, : \, | \langle s,y \rangle| \le 1 \textup{ for all } s \in \D \} $.
\begin{proof}[Proof of Proposition \ref{Prop:ExtremePointsMD}]
	Let $ y \in M_\D $. First, we assume that the set
	\[
	A_y = \{ s \in \D\, : \,|\langle s,y \rangle  | = 1  \}
	\]
	does not span $ \Cd $.
	Then, there exists $ u \in \Cd $ being orthogonal to all elements in $ A_y $ and, since $ d $ is finite, we can find $ \varepsilon > 0 $ such that 
	$ y \pm \varepsilon u \in M_\D $ and $ y = \frac{1}{2} ((y+\varepsilon u)+ (y - \varepsilon u)) $ is a proper convex combination of $ y \pm \varepsilon $. Hence, $ y $ is not an extreme point of $ M_\D $.
	
	Conversely, assume that $ A_y $ spans $ \Cd $  and that $ y = \alpha u + (1-\alpha)v $ for some $ u,v \in M_\D $.
	For every $ s \in A_y $ there is $ \phi_s \in \R $ such that 
	\begin{align*}
		1 = e^{i\phi_s}\langle s,y \rangle   = \alpha e^{i\phi_s} \langle s,u \rangle   + (1-\alpha) e^{i\phi_s} \langle s,v \rangle  ,
	\end{align*}
	hence, 
	\begin{align*}
		\left(e^{i \phi_s}  \langle s,u \rangle \right)^R =  \left(e^{i \phi_s}  \langle s,v \rangle \right)^R = 1.
	\end{align*}
	But as $ | \langle s,u \rangle   | \le 1 $ and $ | \langle s,v \rangle  |  \le 1 $, it must hold that 
	\begin{align*}
		\left(e^{i \phi_s} \langle s,u \rangle  \right)^I =  \left(e^{i \phi_s} \langle s,v \rangle  \right)^I  = 0 .
	\end{align*}
	Since the elements of $ A_y $ span $ \Cd $, the system 
	\begin{align*}
		e^{i \phi_s} \langle s,w \rangle  = 1 \textup{ for all } s \in A_y \textup{ and } w \in \Cd
	\end{align*}
	has the unique solution $ y $, so $ y = u = v $ and $ y $ is  an extreme point of $ M_y $.
\end{proof}

We will continue with the proof of Lemma \ref{Lemma:CS}, which is a consequence of complementary slackness.
\begin{proof}[Proof of Lemma \ref{Lemma:CS}]
	If $(c_s,t_s)_{s \in \D}$ is optimal for the primal and $ (y,(z_s)_{s \in \D}) $ optimal for the dual, then complementary slackness \cite{Alizadeh2003} enforces
	\begin{align*}
		\sum_{s \in \D } (c_s,t_s) \cdot z_s = 0,
	\end{align*}
	but as we have 
	\begin{align*}
		z_s =(- \langle s,y \rangle ^R, -\langle s,y \rangle ^I, 1), 
	\end{align*}
	due to the duality constraint \eqref{FirstDualConstraint}, we can rewrite this as
	\begin{align*}
		\sum_{s \in \D } t_s = \sum_{s \in \D } c_s^R \langle s,y \rangle ^R + c_s^I \langle s,y \rangle ^I. 
	\end{align*}
	Applying Cauchy-Schwarz to each term of the the right hand side we obtain
	\begin{align*}
		\sum_{s \in \D } c_s^R \langle s,y \rangle ^R + c_s^I
		\langle s,y \rangle ^I &\;\le\; \sum_{s \in \D } \norm {(c_s^R, c_s^I)}_2 \cdot  \norm{\, \big(\langle s,y \rangle ^R, \langle s,y \rangle ^I\big) \,}_2 \\
		& \; = \; \sum_{s \in \D }  \norm {c_s}_2  \cdot |\langle s,y \rangle |   \\
		& \; \le \; \sum_{s \in \D } t_s,
	\end{align*}
	where the last inequality follows from $ (c_s,t_s) \in \mathcal{L}^{2+1} $ and $ |\langle s,y \rangle | \le 1 $ for all $ s \in \D $.
	Consequently, we have equality in each step. This leads to the conditions given in the lemma because:
	\begin{enumerate}
		\item [(I)] If $ c_s \neq 0 $, then $ |\langle s,y \rangle | = 1 $, but by the first inequality the vector $ (c_s^R, c_s^I) $ must be proportional to $ (\langle s,y \rangle ^R, \langle s,y \rangle ^I) $, hence 
		$ \langle s,y \rangle = \frac{c_s}{|c_s|} $.
		\item [(II)] If $ |\langle s,y \rangle| < 1 $, then $ c_s = 0 $. \qedhere
	\end{enumerate}
\end{proof}

\bibliography{References}

\end{document}